\documentclass[10pt,preprintnumbers,aps,amssymb,nofootinbib,amsmath]
{revtex4}
\usepackage{epsfig,epsf}
\usepackage{graphicx}
\usepackage{verbatim}
\usepackage{epstopdf}
\usepackage{bm} 
\newcommand{\beq}{\begin{equation}}
\newcommand{\beql}[1]{\begin{equation}\label{#1}}
\newcommand{\eeq}{\end{equation}}
\newcommand{\bea}{\begin{eqnarray}}
\newcommand{\eea}{\end{eqnarray}}
\newcommand{\be}{\begin{eqnarray}}
\newcommand{\ee}{\end{eqnarray}}
\def\eq#1{{(\ref{#1})}}
\def\fig#1{{Fig.~\ref{#1}}}

\newcommand{\as}{\alpha_s}

\newcommand{\Lb}{\left(}
\newcommand{\Rb}{\right)}

\newcommand{\un}{\underline}
\begin{document}

\preprint{BNL-NT, TAUP-2882-08, RBRC-755}


\title{Gluon saturation effects on J/$\mathbf{\Psi}$ production in heavy ion collisions}

\author{Dmitri Kharzeev$^a$, Eugene Levin$^b$, Marzia Nardi$^c$ and
Kirill Tuchin$^{d,e}$\\}

\affiliation{
$^a\,$Department of Physics, Brookhaven National Laboratory,
Upton, New York 11973-5000, USA\\
$^b\,$HEP Department, School of Physics,
Raymond and Beverly Sackler Faculty of Exact Science,
Tel Aviv University, Tel Aviv 69978, Israel\\
$^c$Istituto   Nazionale  di Fisica   Nucleare, Sezione  di Torino, 
via P.Giuria 1, I-10125 Torino, Italy \\
$^d\,$Department of Physics and Astronomy, Iowa State University, Ames, IA 50011\\
$^e\,$RIKEN BNL Research Center, Upton, NY 11973-5000\\}

\date{\today}


\begin{abstract}

We consider a novel mechanism for $J/\Psi$ production in nuclear collisions arising due to the high density of gluons. The resulting $J/\Psi$ production cross section is evaluated as a function of rapidity and centrality. We compute the nuclear modification factor and show that the rapidity distribution of the produced $J/\Psi$'s is significantly more narrow in $AA$ collisions due to the gluon saturation effects. Our results indicate that gluon saturation in the 
colliding nuclei is a significant source of $J/\Psi$ suppression and can explain the experimentally observed rapidity and centrality dependencies of the effect.

\end{abstract}

\maketitle

\section{Introduction}

\begin{figure}[ht]
      \includegraphics[width=10cm]{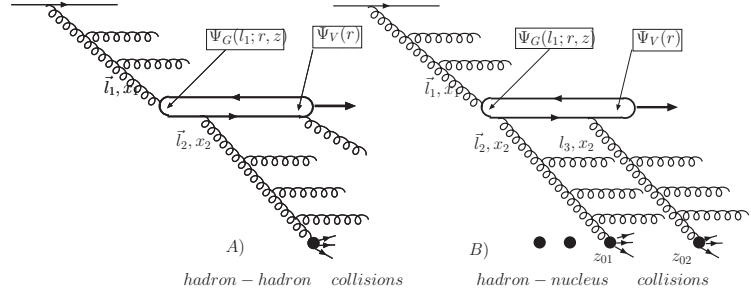} 
\caption{The  process of inclusive J/$\Psi$  production in
  hadron-hadron (\protect\fig{psi1}-A) and  in hadron-nucleus
  collisions (\protect\fig{psi1}-B).  }
\label{psi1}
\end{figure}

The mechanism of $J/\Psi$ production in high energy nuclear collisions can be expected to differ from that in hadron-hadron collisions. Consider first the $J/\Psi$ production in
hadron--hadron collisions. The leading contribution is given by the
two-gluon fusion (i) $G+G\to J/\Psi + \,\mathrm{soft\,\, gluon}$, see \fig{psi1}-A. This process is of the order $\mathcal{O}(\as^5)$. The three-gluon fusion (ii) $G+G+G\to J/\Psi$, see \fig{psi1}-B, is parametrically suppressed as it is proportional to
$\mathcal{O}(\as^6)$. However, in hadron-nucleus collisions an
additional gluon can be attached to the nucleus. This brings in an
additional enhancement by a factor $\sim A^{1/3}$. If the collision energy is high enough, the coherence length becomes much larger than the size of the interaction region. In this case all $A$ nucleons interact coherently as a quasi-classical field \cite{MV,JIMWLK,Kovchegov:1999yj,Kovchegov:1996ty}. In the quasi-classical 
approximation $\as^2A^{1/3}\sim 1$. Therefore, the three-gluon fusion is actually \emph{enhanced} by $1/\as$ as compared to the two-gluon fusion process. Similar conclusion holds for heavy-ion collisions (we do not consider any final state processes  leading to a possible formation of the Quark-Gluon Plasma).

In this letter we calculate the rapidity and centrality dependence of $J/\Psi$ production in $AA$ collisions taking into account the gluon saturation effects \cite{GLR,MUQI}. The case of $J/\Psi$ production in $dAu$ collisions was considered by two of us some time ago \cite{KT} where exactly the same mechanism was discussed, though the nuclear geometry was oversimplified. We note a reasonable agreement of this earlier approach with the $dA$ data; a more definite conclusion can be reached once the higher statistics data becomes available in the near future. Our goal is to understand to what extent the cold nuclear matter effects are responsible for the $J/\Psi$ suppression in $AA$ collisions. In our calculation we use the dipole model \cite{dipole} and take into account realistic nuclear geometry. In Ref.~\cite{KT}  detailed arguments were given which justify the application of the dipole model to the calculation of $J/\Psi$ production
at  RHIC. It was argued that (i) the coherence length for the production of the $c\bar c$ pair is sufficiently larger than the longitudinal extent of the interaction region. This
means that the development of the light-cone ``wave function'' happens
a long time before the collision.  (ii) Formation of $J/\Psi$ is characterized by a time scale on the order of the inverse binding energy. This time is certainly much larger than the $c\bar c$ production 
coherence length (by a factor of $\sim 1/\as^2$) implying that the
formation process takes place far away from the
nucleus. Therefore, in the following, we concentrate on the dynamics
of $c \bar{c}$ pair interactions with the nucleus.

\section{New mechanism of $J/\Psi$ production}


A particularly helpful insight into the nature of the new production mechanism, depicted in \fig{psi1}-B 
is obtained   if we note that the three-gluon contribution 
is suppressed as compared to the two-gluon fusion mechanism (\fig{psi1}-A) by an
additional factor $r^2$, where $r$ is $c\bar c$ transverse separation such that $(2m_c)^{-1}\lesssim r\lesssim (2m_c\as)^{-1}$.  This
factor arises since  we need to have three gluons in the area  of the
order of $ r^2$.  In other words, it means that this reaction is
originated from the next-twist contribution. However,  in the hadron -
nucleus interactions the next-twist contribution appears always in the
dimensionless  combination $r^2Q_s^2$ with the saturation scale
$Q_s$. The saturation scale is proportional to $A^{1/3}$ which
compensates for the smallness of $r$.  
The dominance of  the higher twist process is the main idea of \cite{KT} 
and we develop it in this paper in the case of heavy-ion collisions. 
  
\fig{psi1}-B represents the contribution of the order $(\as^2 A^{1/3})^2$. 
In general,   there must be an odd number of gluons connected to the charm fermion line because the quantum numbers  of $J/\Psi$ and of gluon are $1^{--}$. Therefore, each inelastic interaction of  the $c \bar{c}$ pair must involve two
nucleons and hence is of the order $(\as^2 A^{1/3})^{2n}$, where $n=1,2,\ldots, A/2 $ is the number of nucleon pairs. To take this into account we write the cross section as  the sum over all inelastic processes (labeled by the index
$n$). This sum  involves only even number of interactions. For a heavy nucleus $A\gg 1$ we have 
\be
 \frac{d \sigma_{in}(pA)}{d Y\, d^2b}& =&  \,\,\, C_F\, x_1
G(x_1,m^2_c) \,\label{PP3}\\ 
&&\times\int^{2 R_A}_0 \rho\,\hat\sigma_{in}(x_2,r,r')\,d\,z_{0}
\,\int \,d^2\,r \,\Psi_G(l_1,r,z=1/2)\,\Psi_V(r)\,\otimes\,\int\,d^2
r' \Psi_G(l_1,r',z=1/2)\,\Psi_V(r')\nonumber \\ 
&& \times  \Lb\,e^{-(\sigma(x_2,r^2)
  \,+\,\sigma(x_2,r'^2))\,\rho\,2\,R_A } \,\sum^{\infty}_{n=0} 
\int^{2 R_A}_{z_0}\,d\,z_{1}\, \int^{2 R_A}_{z_1} d z_2 \dots\int^{2
  R_A}_{z_{2n}} d z_{2n +1} \,\rho^{2n+1}\, 
\hat\sigma^{2n+1}_{in}(x_2,r,r')\Rb \nonumber
\ee
where  $\Psi_G \otimes \Psi_V$ is the projection of the $J/\Psi$ light-cone
``wave-function'' onto the virtual gluon one  \cite{KT,XS}, $z_i$'s are the nucleon longitudinal coordinates in the nucleus and 
\beq \label{XSIN}
\hat\sigma_{in}(x_2,r,r')\,\,\equiv\,\,
\sigma(x_2, r^2) + \sigma(x_2, r'^2)\,-\,\sigma(x_2,(\un{r} - \un{r}')^2)\,.
\eeq

After integration over $z_i$'s  and summation over $n$ we obtain the
following formula 
 \be
 \frac{d \sigma_{in}(pA)}{d Y\, d^2b} &= & C_F\,\, x_1
 G(x_1,m^2_c) \, \label{PP5} 
 \int d^2\,r
\Psi_G(l_1,r,z=1/2)\,\Psi_V(r)\otimes\int\,d^2 r'
\Psi_G(l_1,r',z=1/2)\,\Psi_V(r') \nonumber \\
&\times&  \frac{1}{2}\bigg\{ \exp\Big( - \sigma\Lb x_2,(\un{r} -
  \un{r}')^2\Rb\,\rho\,2R_A\Big) \bigg. 
 \bigg.+ \exp\Big( -
( \sigma(x_2,r) + \sigma(x_2,r') +\hat
\sigma_{in}(x_2,r,r'))\,\rho\,2R_A\Big) \nonumber \\ && 
\hspace{5cm}
-2\,\exp\Big( -
(\sigma(x_2,r) + \sigma(x_2,r'))\,\rho\,2 R_A\Big) \bigg\}~.
 \ee
The color factor in \eq{PP3} and \eq{PP5} is calculated in the $N_c\gg 1$ limit.  
  
In the quasi-classical approximation the  gluon saturation is given by
\cite{dipole}
$ Q^2_{s,A}( x)\,=\,4\,\pi^2\as^2\,\rho\,T(\underline{b})$,
 where $\rho$ is the nucleon density in a nucleus, $N_c$ is the number
 of colours, $\un b$ is the impact parameter and $T(b)$ is the optical
 width of the nucleus.  $Q_{s,A}$  determines the scale of the
 typical transverse momenta  for inclusive gluon production
 \cite{KOTU}. Its value was extracted from the fit   
 of the multiplicities  of nuclear reactions at RHIC \cite{KN1,KLN} and 
 from fits of the  $F_2$ structure function in DIS
 \cite{MOD,MOD1,MOD2,MOD3}.


 To generalize  \eq{PP5} to the case of $AA$ collisions we use the approach suggested by Kovchegov \cite{Kovchegov:2000hz} and replace the lowest order gluon field correlation function by the full MV formula as follows 
 \beq \label{RPLCE1}
\frac{\as \pi^2}{3} x_1G(x_1,m^2_c)\,\rightarrow\,\frac{d^2b}{r^2}\,\Lb
1\, -\, \exp \Lb - \frac{r^2\,Q^2_{s,A_1}}{8} \Rb  \Rb \,.
\eeq
Furthermore, noting that  the dominant contribution to the integral over $r'$ in \eq{PP5} comes from the region $r\gg r'$ \cite{KT} we obtain
 \beq
\frac{1}{S_A}\frac{d \sigma(AA)}{d Y\, d^2b} \propto
 Q^2_{s,A_1}(x_1)\,Q^2_{s,A_2}(x_2)\,(Q^2_{s,A_1}(x_1)+
 Q^2_{s,A_2}(x_2))\,  \int_0^\infty d \zeta\, \zeta^9
 \,K_2(\zeta)\,e^{- \frac{\zeta^2}{8\,m^2_c}\,\Lb
   Q^2_{s,A_1}(x_1)\,\,+\,\,Q^2_{s,A_2}(x_2)\Rb }\,. \label{IP4} 
 \eeq
Eq.~\eq{IP4} is derived in the quasi-classical approximation which takes into account multiple scattering of the $c\bar c$ pair in the cold nuclear medium. At forward rapidities at RHIC and at LHC the gluon distribution functions \eq{RPLCE1} evolve according to the evolution equations of the color glass condensate. Inclusion of this evolution in the case of $J/\Psi$ production presents a formidable technical challenge. Therefore we adopt a  phenomenological approach of \cite{KLN}  in which the quantum evolution is encoded in the energy/rapidity dependence of the saturation scales.

In our numerical calculations we take explicit account of the impact parameter dependence of the saturation scales of each nucleus. We employ the Glauber  approximation and assume that the nucleons are small  compared to the
size of the nuclei. The number of $J/\Psi$'s inclusively produced in ion--ion
collisions at given rapidity $Y$ and centrality characterized by
$b$ reads 
 \be
 \frac{d N^{A A}(Y,b)}{ d Y}&=& C \frac{d N^{pp}(Y)}{d Y}\,\,\int d^2
 s\, \,T_{A_1}(\un s)\,T_{A_2} \Lb\un{b} - \un{s}\Rb\, 
\Lb  Q^2_{s,A_1}\Lb x_1,\un{s}\Rb \,\,+\,\,Q^2_{s,A_2}\Lb x_2,\un{b} -
\un{s}\Rb\Rb \,\frac{1}{m^2_c} 
 \nonumber\\
 & &
 \times  \int^{\infty}_0\!\! d \zeta\,\,\zeta^9 \,\,K_2(\zeta)\,
 \,\,\exp\Lb - \frac{\zeta^2}{8 m^2_c}\,\Lb
 Q^2_{s,A_1}(x_1,\un{s})\,+\,Q^2_{s,A_2}(x_2, \un{b} - \un{s})\Rb 
 \Rb\,.\label{IP5}
  \ee
   where $x_{1/2}=\frac{m_{J/\Psi, t}}{\sqrt{s}}\,e^{\mp Y}$   
with $m^2_{J/\Psi, t}=m^2_{J/\Psi}+p^2_t$,  $p_t$ being the transverse momentum  $J/\Psi$.
 The overall normalization constant $C$  includes the color  and the
geometric factors $C^2_F/(4 \pi^2 \as S_p)$ where $S_p$ 
is interaction area in proton--proton collisions. $C$ also  includes
the amplitude of quark--antiquark annihilation into $J/\Psi$ and a
soft gluon in the case of $pp$ collisions. This
amplitude as well as the  mechanism of \fig{psi1}-A  have a significant theoretical uncertainty.  
Therefore, we decided to parameterize these contributions by an
overall normalization constant in  \eq{IP5}.

\begin{figure}[ht]
\begin{tabular}{cc}
      \includegraphics[width=8cm]{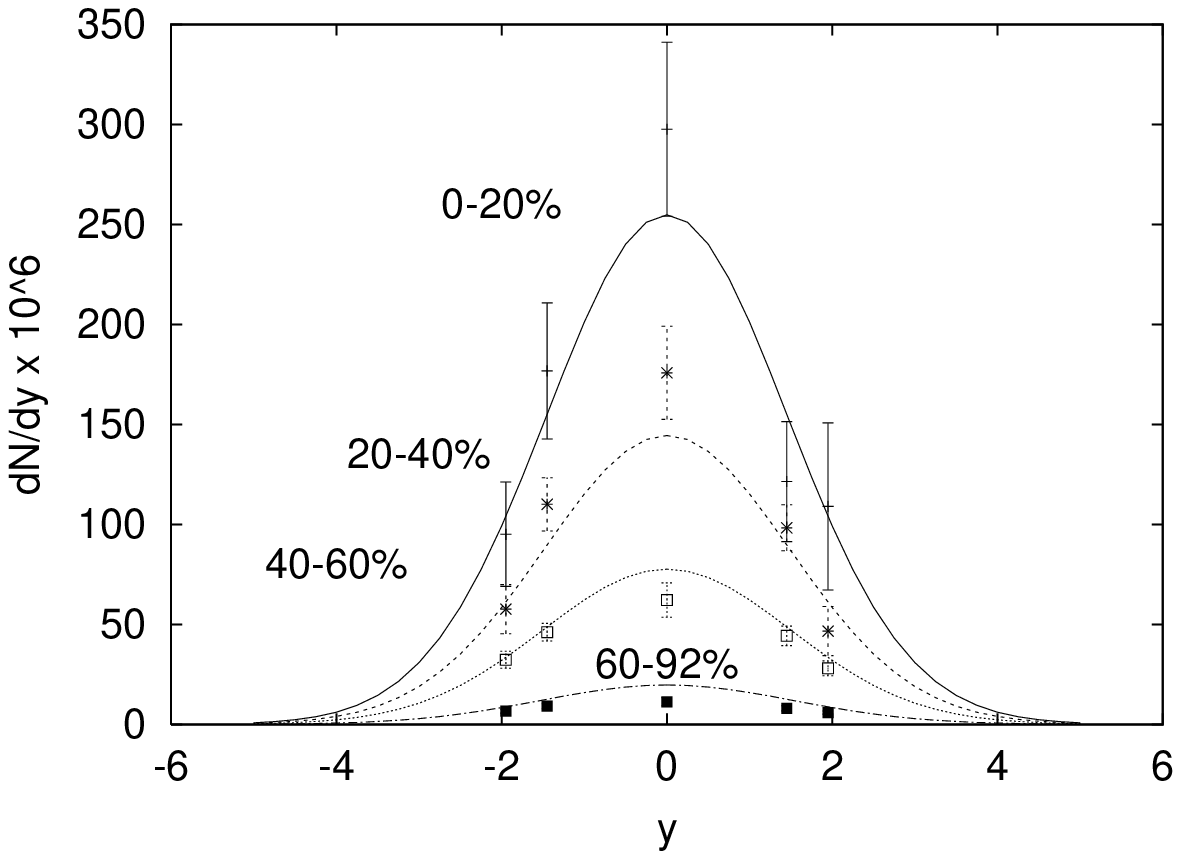} &
      \includegraphics[width=8cm]{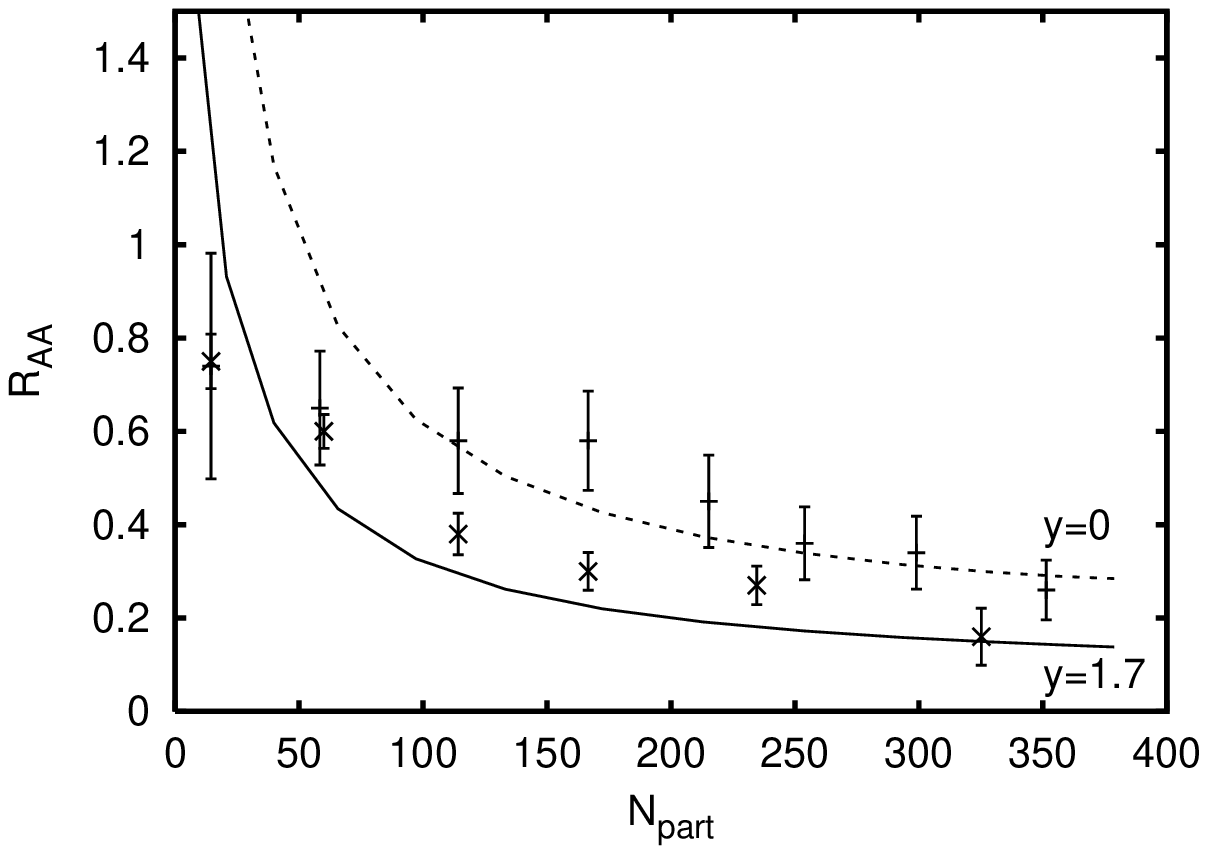}\\
      (a) & (b)
 \end{tabular}     
\caption{(a) $J/\psi$ rapidity distribution in Au-Au collisions for
  different centrality cuts. (b) Nuclear modification factor for $J/\Psi$ production in heavy-ion collisions for different rapidities. Experimental data from \cite{Adare:2006ns}.}
\label{fig:BdNdy}
\end{figure}

The  rapidity distribution of $J/\psi$'s in $pp$ collisions, the
factor ${d N^{pp}}/{d Y}$ appearing in Eq. \ref{IP5}, is fitted to
the experimental data given in \cite{Adare:2006kf} with a single
gaussian. In figure \ref{fig:BdNdy}(a)
the results provided by Eq.~\eq{IP5} are then compared
to the experimental data from PHENIX Collaboration \cite{Adare:2006ns} for
Au-Au collisions at $\sqrt{s}=200$ GeV. The global normalization factor $C$ is
found from the overall fit. There are no other free
parameters.  The agreement of the theoretical results
with experimental data is reasonable. It is evident that the effect of the gluon saturation on the  $J/\psi$ rapidity distribution in nucleus-nucleus collisions is to make its width a decreasing function of centrality. The distribution in the most central bin is significantly  more narrow than in the  peripheral bin.

It is important that we describe well the data in the semi-peripheral region. This ensures that our model gives a good description of the $J/\Psi$ production in $dAu$ collisions. We also note that an earlier approach \cite{KT} in which the same model was employed (albeit with an oversimplified nuclear geometry) provided a reasonable description of the data.  Still a more detailed investigation is required which takes into account the exact deuteron and gold nuclear distributions. This will allow a model-independent fixing of the overall normalization constant $C$. Such an analysis will be presented in the near future. 

Previously, the initial-state effects on the nuclear suppression of $J/\psi $'s have been estimated \cite{Karsch:2005nk,Granier de Cassagnac:2007aj} through the product of nuclear modification factors measured in $dAu$ collisions. This estimate holds as long as the collinear 
factorization holds for the process of J/Psi production (this would be true for example 
in the limit of small cc-bar dipole size). Here we find that  the effects beyond the 
collinear factorization are strong, and may account for a large part of the 
suppression measured in Au-Au, especially at forward rapidities (where the density 
of gluons in the initial state, inside one of the colliding nuclei, is the largest).  

To emphasize the nuclear dependence of the inclusive cross sections it is convenient to introduce the nuclear modification factor
\beq\label{nmf}
R_{AA}(y,N_\mathrm{part})=\frac{ \frac{dN^{AA}}{dy} }
{N_\mathrm{coll}\,\frac{dN^{pp}}{dy}}\,.
\eeq
It is normalized in such a way that no nuclear effect would correspond to $R_{AA}=1$. In \fig{fig:BdNdy}(b) we plot the result of our calculation. The nuclear modification factor exhibits the following  two important features: (i) unlike the open charm production, $J/\Psi$ is suppressed even at $y=0$;  (ii) cold nuclear matter effects account for a significant part of the ``anomalous" $J/\Psi$ suppression in heavy-ion collisions both at $y=0$ and $y=2$.


\section{Conclusions}\label{concl}

The main results of this paper are exhibited in Fig.~\ref{fig:BdNdy}. 
It is seen that the rapidity and centrality dependence of $J/\Psi$ production are reproduced with a reasonable accuracy even without taking into account any hot nuclear medium effects. 
This observation allows to conclude that a fair amount (and perhaps most) of the $J/\Psi$ suppression in high energy heavy-ion collisions arises from the cold nuclear matter effects. In other words, $J/\Psi$ is expected to be strongly suppressed even if there were no hot nuclear matter effect produced. In this sense we can talk about the separation of the cold and hot nuclear medium effects as advertised in the Introduction.

A certain fraction of $J/\Psi$ suppression in forward direction can be attributed to 
suppression of the constituent $c$ and $\bar c$ quarks \cite{Tuchin:2004rb,Kovchegov:2006qn,Gelis:2003vh,Blaizot:2004wv} which would lead to suppression of $D$-mesons \cite{Tuchin:2007pf}. However, $D$-mesons are not predicted to be suppressed at central rapidities \cite{Tuchin:2007pf}. 

The reason for $J/\Psi$ suppression at mid-rapidities is that the multiple scattering of $c\bar c$ in the cold nuclear medium increases the relative momentum between the quark and antiquark, which makes the bound state formation less probable. It was proven in \cite{KKT} that unless quantum $\log(1/x)$ corrections become important, the inclusive gluon production satisfies the sum rule that requires the nuclear modification factor to be of order unity. Similar sum rule holds for heavy quark production but fails in the case of a 
bound states, such as $J/\Psi$.

We realize that although our calculation gives the parametrically leading result at high gluon density, other production channels involving the gluon radiation in the final state and the color octet mechanism of  $J/\Psi$ production may give phenomenologically significant contributions. These are likely to become the leading mechanisms in the peripheral collisions where the strength of the gluon fields is significantly diminished. However, we believe that our main results are robust for central high-energy collisions of heavy ions. These results imply that the observed $J/\Psi$ suppression is a result of  an interplay between the cold and hot nuclear matter effects.

\vskip0.3cm
{\bf Acknowledgments.}
The work of  D.K. was supported by the U.S. Department of Energy under Contract No. DE-AC02-98CH10886.
K.T. was supported in part by the U.S. Department of Energy under Grant No. DE-FG02-87ER40371; he would like to
thank RIKEN, BNL,
and the U.S. Department of Energy (Contract No. DE-AC02-98CH10886) for providing facilities essential
for the completion of this work.  This research of E.L.  was supported in part by  a
grant from Ministry of Science, Culture \& Sport, Israel \& the
Russian  Foundation for Basic research of  the Russian Federation,
and by BSF grant \# 20004019.
\vskip0.3cm


\end{document}